# A Ring Router Microarchitecture for NoCs

Wo-Tak Wu

**Abstract**— Network-on-Chip (NoC) has become a popular choice for connecting a large number of processing cores in chip multiprocessor design. In a conventional NoC design, most of the area in the router is occupied by the buffers and the crossbar switch. These two components also consume the majority of the router's power. Much of the research in NoC has been based on the conventional router microarchitecture. We propose a novel router microarchitecture that treats the router itself as a small network of the ring topology. It eliminates the large crossbar switch in the conventional design. In addition, network latency is much reduced. Simulation and circuit synthesis show that the proposed microarchitecture can reduce the latency, area and power by 53%, 34% and 27%, respectively, compared to the conventional design.

—————————— ◆ ——————————

## 1 INTRODUCTION

With advances in semiconductor process technology in recent decades, the number of processing cores that can be fabricated on a single die has been increasing steadily. Nowadays, we even have chip multiprocessors (CMP) with a thousand or more cores [1]. An effective way of connecting a large number of cores is using the NoC approach. Since the cores are typically laid out in a regular pattern on a die, a NoC with a rectangular mesh topology is a sensible design choice. Fig. 1 shows an example of a 4x4 mesh network with 16 cores, each connecting to the network through a router.

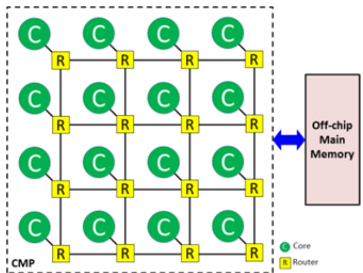

Fig. 1. System configuration.

The conventional router microarchitecture [2] is shown in Fig. a. The router has five inputs and five outputs for connections with the core and four neighboring routers. The input units hold the incoming packets in a set of buffers. The route computation unit determines where the packets will leave the router. The allocators control the flow of packets through the crossbar switch.

It is well known that the area and power of a router are dominated by the buffers and the crossbar switch [3]. In this work, we propose a novel router microarchitecture that is tailored for 2D mesh networks, which is amicable for the prevailing planar chip layout. The proposed design is basically a small network inside the router. It adopts a ring topology as shown in Fig. 1a. Each input-output pair at the router interface is directly connected to an exchange (XC) in the router. The five exchanges are

————————————————
- *The author is with the Department of Electrical and Computer Engineering, The University of Arizona, Tucson, AZ 85721. E-mail: wotakwu@email.arizona.edu*

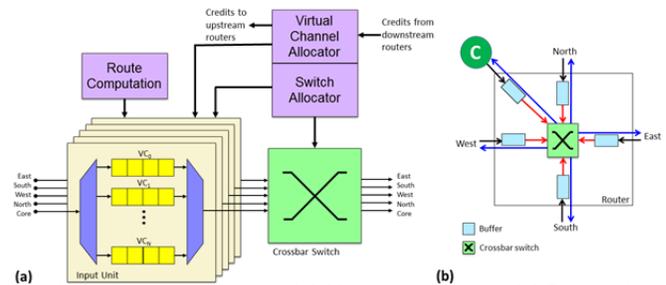

Fig. 2. Conventional router: (a) Microarchitecture; (b) Datapath.

connected in a ring manner. The arrangement of exchanges is optimized for the popular dimension-order routing (DOR) algorithm [2] such that packets mostly traverse only two exchanges in the router.

Compared to the conventional router design, the proposed microarchitecture eliminates the crossbar switch, which would save much area and reduce power consumption. In addition, it takes only one cycle to traverse an exchange. A packet typically takes two cycles to go through the router. As a result, the proposed microarchitecture also reduces network latency.

We use a cycle-accurate network simulator to validate the proposed design on an 8x8 mesh network. We also use commercial CAD tools to estimate accurately the router's area and power, and to validate the functionality of the circuitry. On running a range of synthetic traffic loads and real applications, even with fewer buffers, the proposed microarchitecture outperforms the conventional design. The network latency is reduced by 53%, the router area by 34%, and the power consumption by 27%.

## 2 CONVENTIONAL ROUTER MICROARCHITECTURE

Fig. b shows the datapath of the conventional router microarchitecture. A packet entering the router is immediately stored in one of the virtual channels (VCs) in the buffer. Then it competes with other pending packets (if any) in other input units and in its own as well to get access to the crossbar switch to leave the router. The size of the crossbar switch grows exponentially with the number of inputs and outputs. With five inputs and five outputs, the crossbar switch takes up a sizable area in the router.

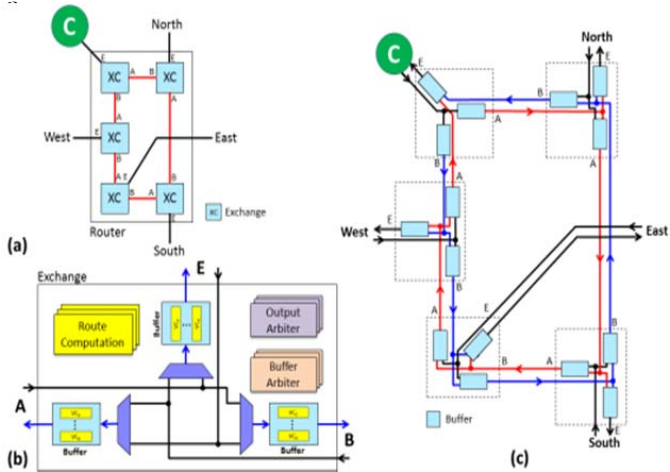

Fig. 1. Ring Router: (a) Ring network; (b) Exchange; (c) Datapath.

In addition to buffering and switch traversal, two resource allocation processes also need to take place. As a result, it takes at least four cycles to go through the router end-to-end. To attain high performance, a four-stage pipeline is constructed to achieve a maximum throughput of one packet per cycle at each output port.

## 3 PROPOSED ROUTER MICROARCHITECTURE

### 3.1 Hardware Design

While the conventional router architecture is designed for various types of networks, the proposed microarchitecture specifically targets 2D mesh networks, which typically do not have a large number of inputs/outputs. It simplifies the router design with a small ring network of five exchanges, four for the neighboring routers and one for the core. The arrangement of exchanges in the ring network minimizes the hop count within the router for DOR, which routes packets in the x-dimension first followed by the y-dimension. Most of the time, packets go through a router either vertically or horizontally. They make a turn when changing dimension is needed, which happens at most once in the entire route. A packet traveling vertically involves only the north and south exchanges, and horizontally, the west and east exchanges.

Each exchange has only three bidirectional ports, as shown in Fig. 1a. Ports A and B connect to the other exchanges in the router, and Port E ('E' stands for external) connects to another router or the core. Fig. 1b shows a block diagram of the exchange design. A packet entering the exchange can only leave from one of the other two ports. Therefore, no loopback is allowed within the exchange. This simplifies the mux design that sits just before the buffer. Unlike in the conventional microarchitecture, there is no large crossbar switch; there are only simple 2:1 muxes. As a result, large chip real estate is saved.

Each port in the exchange is associated with a buffer, which has the same structure as that of the conventional design. As seen in Fig. 1b, there are three types of control units in exchange. The route computation unit determines a packet's exit port in an exchange based on the destination node coordinates carried in the packet. The output arbiter selects which virtual channel in the buffer outputs a packet to the exchange downstream. The buffer arbiter determines which one of the two inputs can write to the associated buffer. For example, buffer arbiter E determines whether entry Port A or B can write to the buffer connecting to exit Port E.

### 3.2 Timing

Before a packet is written into a buffer, the route computation unit in the current exchange determines which exit port to take in the exchange downstream. A write request is then immediately made to the buffer arbiter in the exchange downstream, hoping that it will be granted in the next clock cycle. Buffers are dual-ported; they have separate read and write ports, which work independently with their own clocks. We use both edges of the clock—rising for read and falling for write. Writing to the buffer happens in the second half of a clock cycle. In the next clock cycle, if granted access from the buffer arbiter downstream, a packet will be output to leave the current exchange. As a result, a packet can traverse an exchange in just one clock cycle.

### 3.3 Routing

Since packet buffering is done at the exit point of an exchange (as opposed to the entry point in the conventional router), routing is done for the exchange downstream—a lookahead routing scheme.

All the links attached to the exchanges are unidirectional. Therefore, there are actually two rings in the network, one in the clockwise direction and the other in the counter-clockwise direction, as shown in Fig. 1c in red and blue lines, respectively. The ring network is actually not of a perfect ring. To avoid cyclic dependency in the ring network, there is a logical disjoint (created by the routing mechanism) at the core exchange. Without such disjoint, deadlock may form in the loop in heavy network traffic conditions. Consequently, only packets injected into the core port can be looped back to the same port in the router. For DOR algorithm, which changes direction at most once during the entire packet route, this disjoint imposes minimal impact on the performance. Additionally, a livelock condition will not happen because packets cannot circulate in the exchange.

The consequence of the disjoint at the core exchange is that the shortest path between the north and west exchanges through the core exchange does not exist. Fortunately, the alternate path takes only one extra hop by going through the east and south exchanges. Since it is the only change of dimension in the route, the impact on latency would be quite small.

Table 1 shows how packets are routed through exchanges inside a router. Each hop takes one cycle. The minimum hop count is two. Most of the hop counts are either two or three. Only the north-west path takes four hops. The loopback from the core takes six hops.

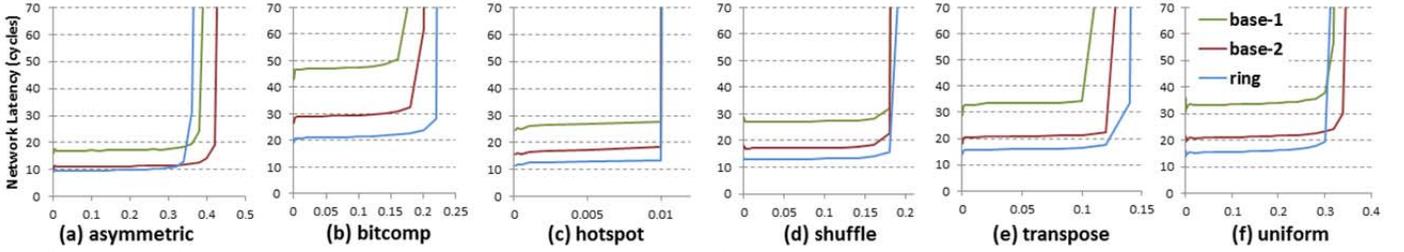

Fig. 2. Network latency performance over a range of synthetic traffic patterns. The horizontal axis is the flit injection rate in flits/cycle.

**Table 1 Routing inside the ring router**

| Source | Destination | Stops | Hops |
|---|---|---|---|
| Core | Core | North, South, East, West | 6 |
| Core | North | none | 2 |
| Core | East | West | 3 |
| Core | South | North | 3 |
| Core | West | none | 2 |
| North | East | South | 3 |
| North | South | none | 2 |
| North | West | South, East | 4 |
| East | South | none | 2 |
| East | West | none | 2 |
| South | West | East | 3 |

A round-robin arbitration scheme is adopted for gaining access to the buffers in the exchange. Thus, starvation to any inputs to the router or exchanges will not occur.

## 4 EVALUATION

We augment the cycle-accurate NoC simulator Booksim [4] with the proposed ring router microarchitecture to measure the average network latency and saturation. The latency includes the queuing delay before the packet is injected into the network. We use an 8x8 mesh network to run six different synthetic traffic patterns (*asymmetric, bitcomp, hotspot, shuffle, transpose, uniform*) and the network traces [5] from PARSEC benchmark suite [6]. We set the packet length to be one flit long for the synthetic traffics. The width of the buffer, link and flit are the same, which is 128 bits. Each VC is 8 flits deep. The routing algorithm is DOR.

In addition to the baseline conventional design, we also add a more advanced router for comparison, which is still based on the conventional design but enhanced with lookahead routing and speculative resource allocation features. Altogether, we have three router designs to evaluate.

We implement the router circuitries with Verilog, based on the conventional design by Becker [7]. We synthesize the router circuitries using Cadence's CAD tool, which also provides accurate estimates of the router's power and area with a 45-nm process technology. A separate simulation tool is used to validate the functionality and timing of the circuits.

For the conventional routers, we adopt eight VCs per buffer per input unit. Since there are five input units, the total number of VCs is 40 (5x8). For the ring router, we use two VCs per buffer per port in an exchange. Since there are five exchanges and each has three buffers, the total number of VCs in a router is 30 (5x3x2).

## 5 RESULTS

We denote the baseline conventional design as *base-1*, the one with the advanced features built on top of *base-1* as *base-2*, and the ring router as *ring*.

Fig. 4 shows the average network latencies under different packet injection rates (or traffic densities) across all synthetic traffic patterns. The results look similar from pattern to pattern. From the y-intercepts, we can see that *ring* has lower no-load latency than *base-1* and *base-2*. In Fig. 5, *ring* also shows better latency performance on the entire PARSEC benchmark suite, with an average reduction of 49%, compared to *base-1*.

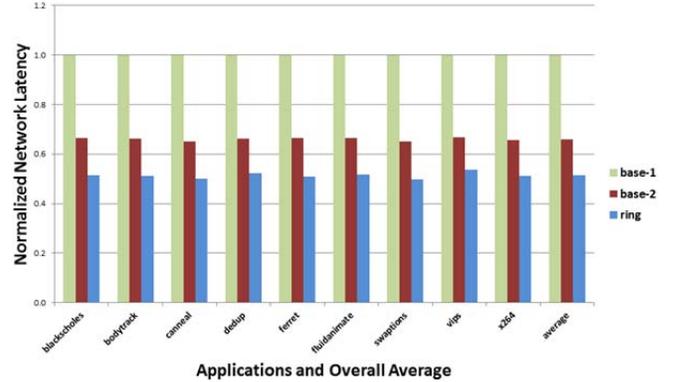

Fig. 5. Network latency performance, normalized to *base-1*, on PARSEC applications.

The sharp rising part of the curves (in Fig. 4) indicates where the network starts to saturate. The saturation point is defined as where the latency is two times or more than the no-load latency. *Ring* saturates at a later point than or at the same point as *base-1* and *base-2* on four out of six traffic patterns: *bitcomp*, *hotspot*, *shuffle* and *transpose*.

Fig. 6 shows the latency and saturation performances averaged over all six synthetic traffic patterns, relative to *base-1*. *Ring* reduces the latency by 53% and performs equally well in saturation. In terms of circuit area and power, they are reduced by 34% and 27%, respectively. *Ring* also outperforms *base-2* by 16%, 35% and 28% in latency, area and power, respectively; however, it does worse by 8% in saturation.

## 6 DISCUSSION

With 25% fewer buffer resources (30 vs. 40 VCs), *ring* clearly outperforms the conventional designs in latency,

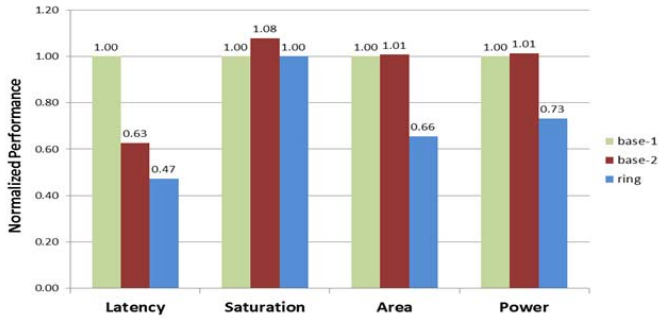

Fig. 6. Four performance metrics normalized to base-1. The lower, the better, except in saturation.

area and power. The elimination of the crossbar switch and adopting fewer buffers clearly help reduce the circuit area and power.

The proposed microarchitecture is a drastically different design from the conventional one. There is no large crossbar switch; buffers are situated at the exit point rather than at the entry point; the router itself is constructed with five identical modules, rather than with many different ones; buffer access arbitration logic can be very simple since it needs to deal with just two inputs. These factors bring about a very simple exchange design, which allows packets to traverse in one clock cycle. As a result, we see a drastic reduction in latency. The highly modular design would also help in chip layout and wiring.

The buffer resource has a big impact on saturation, for more buffer space increases network capacity, thus delaying the onset of saturation. Even with 25% fewer VCs, *ring* performs at the same level as *base-1*, and it is only slightly worse than *base-2*. This performance can be attributed to the architectural design differences that largely offset the big difference in the buffer. With the ring topology in the router, the packet movement is much more dynamic than in the conventional router. Packets go from buffer to buffer when traversing in the router, thus freeing up buffer space constantly, creating more opportunities for new packets to enter the network or router. In other words, the traffic load is more spread out among exchanges in the router.

The proposed microarchitecture provides additional path diversity in the router, which is absent in the conventional architecture. The injection of a packet into the core exchange can take on either the clockwise or counter-clockwise path. If the disjoint at the core exchange can be eliminated, such path diversity is then available to all entry ports to the router. This will facilitate adaptive routing algorithms and fault tolerance.

Simple round-robin algorithm is being used in arbitration. More sophisticated algorithms can probably be adopted. This should help push further out the saturation point.

This is just an initial effort on the proposed microarchitecture. More sophisticated and popular schemes, like datapath by-passing logic and shared buffer management, can be considered to further improve the performance.

## 7 RELATED WORK

Much of the NoC research has been on routing algorithms, buffer design/management, and the underlying router microarchitecture is based on the conventional design. There really has not been much work done on drastic router microarchitecture redesign. Kim [3] partitions the router into two parts, one for each dimension. The crossbar switch still remains, however. His work does provide inspiration on how to arrange the exchanges in our work. Abad et al. [8] also adopt a ring topology in their router. Their design does not use any virtual channels and relies on circulating packets if they cannot exit the router immediately, which could result in substantial power consumption.

## 8 CONCLUSION

We propose a novel router microarchitecture that adopts a ring-topology network in the router. We believe it is a much simpler design than the conventional one. It targets the prevailing 2D mesh networks. The exchanges (network nodes) in the network are of identical and simple design. A packet can traverse the exchange in one clock cycle. The arrangement of exchanges in the ring network is optimized for the prevailing dimension-order-routing algorithm for mesh networks. Simulation and circuit synthesis show superior performance over the conventional router microarchitecture in latency, area and power. More enhancements can be made to push the performance further.